\documentclass[12pt]{article}
\usepackage{graphicx}
\usepackage{amssymb}
\def\pk{(k)}
\def\calg{{\mathcal G}}
\def\gkab{\calg^{\pk a}{}_b}

\def\call{{\mathcal L}}
\def\calj{{\mathcal J}}
\def\calr{{\mathcal R}}
\def\rkab{\calr^{\pk a}{}_b}
\def\gab{g^{a}{}_b}
\def\half{{1\over 2}}

\def\kbar{{\bar k}}
\def\callk{\call_k}
\def\bkab{B^{(k)ab}}
\def\jka{\calj^{(k)a}}
\def\okab{\omega^{(k)ab}}

\def\8pig{8\pi G}

\def\khat{{\hat k}}

\begin{document}

\begin{titlepage}
\vfill
\begin{flushright}
%\today
\end{flushright}

\vfill
%\vskip 1.0cm
\begin{center}
\baselineskip=16pt
{\Large\bf Komar Integrals in Higher (and Lower)}
\vskip 0.5cm
{\Large\bf  Derivative Gravity}
\vskip 0.5cm
{\large {\sl }}
\vskip 10.mm
{\bf David Kastor} \\
%\\[2mm]
\vskip 0.5cm
%\vfill
{

       Department of Physics\\
       University of Massachusetts\\
       Amherst, MA 01003\\
       \texttt{kastor@physics.umass.edu}\\
}
\vspace{6pt}
\end{center}
\vskip 0.5in
\par
\begin{center}
{\bf Abstract}
 \end{center}
\begin{quote}
The Komar integral relation of Einstein gravity is generalized to Lovelock theories of gravity.  This includes, in particular, a new boundary integral for the Komar mass in Einstein gravity with a nonzero cosmological constant, which  has a finite result for asymptotically AdS black holes, without the need for an infinite background subtraction.  Explicit computations of the Komar mass are given for black holes in pure Lovelock gravities of all orders and in general Gauss-Bonnet theories.
\vfill
% \hrule width 5.cm
\vskip 2.mm
\end{quote}
\end{titlepage}

\section{Introduction}
In this paper, we ask whether  Komar integral relations  can be found in  higher  and lower derivative theories of gravity.  By `lower derivative gravity,' we simply mean including a non-zero cosmological constant, corresponding to a zero ({\it i.e.} lower) derivative term in the gravitational action.  
Recall first how  the Komar construction works in Einstein gravity with vanishing cosmological constant \cite{Komar:1958wp}.  Assume that we have an asymptotically flat spacetime with a Killing vector $\xi^a$.  Let $\Sigma$ be a spatial hypersurface with boundary $\partial\Sigma$.  The  Komar boundary integral is defined to be
\begin{equation}\label{komarcharge}
Q=\int_{\partial\Sigma} dS_{ab}\nabla^a\xi^b
\end{equation}
This can be rewritten,  using Gauss' theorem 
$\int_{\partial\Sigma} dS_{ab}\omega^{ab}=\int_{\Sigma} dS_{a}\nabla_b\omega^{ab}$
and the property of Killing vectors $\nabla_a\nabla^a\xi^b=-R^b{}_c\xi^c$ as the volume integral
\begin{equation}\label{volumecharge}
Q=\int_\Sigma dS_a R^a{}_b\, \xi^b.
\end{equation}
If the  vacuum Einstein equations $R^a{}_b=0$ are satisfied, then the result that the original Komar boundary integral must vanish.  We will call this result, that   $Q=0$, a Komar integral relation.

In a black hole spacetime, the hypersurface $\Sigma$ can be taken to have two boundaries, one at infinity and one at the black hole horizon.  The statement $Q=0$ relates the integrals over these two disconnected components of $\partial\Sigma$.  For static black holes, taking $\xi^a$ to be the time translation Killing vector, the boundary term at infinity is proportional to the mass $M$.  The boundary term at the horizon turns out to be proportional to $\kappa A$, where $\kappa$ is the surface gravity and $A$ is the horizon area.  The statement $Q=0$ relating these two boundary integrals then yields the Smarr formula
\begin{equation}\label{smarr}
M = \left( {D-2\over D-3}\right ) {\kappa A\over 8\pi G}
\end{equation}
One of the goals of the present work is to understand whether black holes in higher (and lower) derivative gravity theories satisfy similar Smarr formulas.

How does the discussion above change, if the gravitational Lagrangian is  changed?  The relation between the boundary and volume integral expressions for $Q$ remains the same, since this only depends on Gauss' theorem.  However, the equations of motion  no longer implies that $Q=0$. 
Hence, there is no longer a Komar integral relation.  The question we will ask below is  whether it is possible to find a new Komar boundary integrand, such that the corresponding volume integral again vanishes by virtue of the equations of motion.  One would then again have a Komar integral relation $Q=0$.

Asking whether the Komar integral construction can be generalized to arbitrary higher (and lower) derivative gravity theories would be an unnecessarily complicated starting point.  Even the apparently simple addition of an $R^2$ interaction to the Lagrangian leads to terms in the equations of motion involving the second  derivatives of the curvature tensor.  The subclass of Lovelock gravity theories \cite{Lovelock:1971yv} offers a simpler starting point, with  equations of motion depending only on the Riemann tensor and not on its derviatives.  In this paper, we will focus on the Komar construction in this limited class of theories.

The paper is structured as follows.  The basic formalism of Lovelock  gravity theories is presented in section (\ref{lovelocksection}).  In section (\ref{purelovelock}) a new Komar boundary integrand  is constructed  for the still more limited sub-class of pure Lovelock gravity theories,  whose dynamics are governed by a single Lovelock term.  In section (\ref{purebh}) this result is applied to the black hole solutions of pure Lovelock gravity.  The Komar construction in the simplest non-pure Lovelock theory, Einstein gravity with a non-zero cosmological constant, is studied in section (\ref{cosmoconstant}) and applied to Schwarzschild AdS black holes in section (\ref{schwarz-ads}).
With these results in hand, in section (\ref{generalsection}) we will see how the Komar construction works in a general Lovelock theory  and apply it to black holes in general Gauss-Bonnet gravity in section (\ref{GBsection}).  Section (\ref{conclusions}) contains some concluding remarks.

\section{Lovelock Gravity}\label{lovelocksection}

The Lagrangian for Lovelock gravity \cite{Lovelock:1971yv} has the form
\begin{equation}\label{lovelocklag}
\call = \sum_{k=0}^{\kbar} c_k\call_k
\end{equation}
where each term $\callk$ denotes a certain scalar combination built out of $k$ powers of the Riemann tensor.  The $k=0$ term is given by $\call_0=1$. Its coefficient $c_0$ is proportional to the cosmological constant.  The $k=1$ term is simply the Einstein term $\call_1=R$, while the term quadratic in the curvature is the so-called Gauss-Bonnet term,
\begin{equation}\label{gauss-bonnet}
\call_2 = R_{abcd}R^{abcd}-4R_{ab}R^{ab}+R^2.
\end{equation}
Each of the terms $\callk$ has the property that in dimension $D=2k$ its integral gives the Euler character.  For $D<2k$ the term $\callk$ vanishes identically.  Hence, the $k$th order Lovelock term contributes to the equations of motion only in dimensions $D\ge 2k+1$.  We can therefore assume that the upper limit of the sum in (\ref{lovelocklag}) satisfies $\kbar\le [(D-1)]/2]$.

An explicit expression for the terms in the Lovelock Lagrangian is
\begin{equation}\label{lovelock}
\mathcal{L}_k= {\frac{1}{2^k}}\sqrt{-g}\, \delta^{a_1\dots a_{k}b_1\dots b_k}_{c_1\dots c_{k}d_1\dots d_k} \, 
R_{a_1b_1} {}^{c_1d_1}\dots R_{a_kb_k} {}^{c_kd_k},
\end{equation}
where the $\delta$ symbol denotes the totally anti-symmetrized product of Kronecker delta functions
\begin{equation}
\delta^{a_1\dots a_n}_{b_1\dots b_n} = n!\delta^{[a_1}_{b_1}\dots\delta^{a_n]}_{b_n}
= n!\delta^{a_1}_{[b_1}\dots\delta^{a_n}_{b_n]}
\end{equation}
Unlike generic higher derivative gravity theories, 
the equations of motion of Lovelock gravity depend only on the curvature tensor and not on its derivatives.  The equations of motion can be written as $\calg^a{}_b\equiv \sum_{k\le\kbar} c_k\calg^{\pk a}{}_b=0$, where
\begin{equation}\label{lovelockmotion}
\mathcal{G}^{(k)a}{}_b = -{1\over 2^{(k+1)}}
\delta^{ac_1d_1...c_kd_k}_{be_1f_1...e_kf_k}\,R_{c_1d_1}{}^{e_1f_1}\,....\,R_{c_kd_k}{}^{e_kf_k}.
\end{equation}
At each order $k$ these tensors satisfy the identity $\nabla_a\gkab =0$ by virtue of the Bianchi identity for the Riemann tensor $\nabla_{[a}R_{bc]}{}^{de}=0$ and the anti-symmetrization built into the $\delta$ symbol.

It will be useful to write the Einstein-like tensors $\gkab$ in the form
\begin{equation}\label{einstein-like}
\gkab = \rkab - \half\gab\callk ,
\end{equation}
where the Ricci-like tensor $\rkab$ is given explicitly by
\begin{equation}
\rkab = {k\over 2^k}
\delta^{c_1d_1...c_kd_k}_{b\, f_1...e_kf_k}\,R_{c_1d_1}{}^{a f_1}\,....\,R_{c_kd_k}{}^{e_kf_k}.
\end{equation}
Taking the trace one finds the useful relation 
\begin{equation}\label{trace}
\calr^{\pk a}{}_a = k\callk
\end{equation}
which plays an analogous role in Lovelock gravity to the basic relation $R^a{}_a = R = \call_1$ in Einstein gravity.

In the next section, we will focus on an even more restricted class of theories, which have been called  pure Lovelock gravities, in which the Lagrangian consists only of a single Lovelock term, {\it i.e.} for $k$th order pure Lovelock gravity the Lagrangian is $\call=c\callk$.  
The pure Lovelock theory with $k=1$ is simply Einstein gravity with zero cosmological constant, while for $k=2$ the Lagrangian is a multiple of the Gauss-Bonnet term (\ref{gauss-bonnet}).
The equation of motion is simply $\gkab =0$.  Using the relation (\ref{trace}), we see that this implies that
\begin{equation}\label{pureequation}
\rkab=0
\end{equation}
or pure Lovelock theories, as in pure Einstein gravity.  Note that because of the factor of $k$ on the right hand side of (\ref{trace}) a similar result does not hold in a general Lovelock theory, {\it i.e.} 
generally $\calr^a{}_b \equiv \sum_k c_k\rkab \neq 0$ for solutions to the equations of motion.
Similarly, except for the pure Lovelock theories, the field equations for a general Lovelock theory do not imply that the Lagrangian vanishes on solutions.

\section{Komar integrals in pure Lovelock gravities}\label{purelovelock}

The Komar construction can be generalized to all  pure Lovelock gravities.  It is simplest to start with  pure Gauss-Bonnet gravity,  with Lagrangian 
$\call=c\call_2$ and equations of motion that can be written as
\begin{equation}\label{gbequation}
\calr^{(2)a}{}_b = 2 (R^{acde}R_{bcde}-2 R^a{}_cR^c{}_b
- 2 R^{ac}_{bd}R^d{}_c +R^a{}_b R) = 0.
\end{equation}
In order to make the Komar construction work in this theory, we need to find a new boundary term 
$B^{ab}$ for pure GB gravity such that $\nabla_a B^{ab}$ vanishes by virtue of equation (\ref{gbequation}).
 
A simple counting argument suggests what new terms to include in $B^{ab}$.
In Einstein gravity, the Komar boundary integrand in (\ref{komarcharge}) has a single derivative acting on the Killing vector.  Converting the boundary integral to a volume integral via Gauss's theorem adds a second derivative.  Two derivatives acting on a Killing vector produce a curvature tensor contracted with the Killing vector.   The equations of motion (\ref{gbequation}) are purely quadratic in the curvature.  In order to get such terms in the volume integrand, we need to start with terms in the boundary integral that are already linear in the curvature tensor, as well as being linear in the first derivative of the Killing vector.

Given that the boundary integrand $B^{ab}$ should be antisymmetric, the most general possibility is a linear combination of three terms, having the form
\begin{equation}
B^{ab}=\alpha R\, \nabla^a \xi^b+\beta (R^a{}_c\nabla^c\xi^b-R^b{}_c\nabla^c\xi^a) +
\gamma R^{ab}{}_{cd}\nabla^c\xi^d,
\end{equation}
where $\alpha$, $\beta$ and $\gamma$ are numerical coefficients.
The Komar boundary integral is $Q=\int dS_{ab}B^{ab}$ and
the corresponding volume integrand is obtained by taking the divergrance of $B^{ab}$, giving
given by
\begin{eqnarray}\label{divergence}
\nabla_aB^{ab}&= &(\alpha\nabla_aR +\beta\nabla_cR^c{}_a)\nabla^a\xi^b +
(-\beta\nabla_a R^b{}_c+\gamma\nabla_d R^{db}{}_{ca})\nabla^c\xi^a \\ \nonumber
&\, & + (-\alpha RR^b{}_e - \beta R^a{}_cR^{cb}{}_{ae}-\beta R^b{}_c R^c{}_e - 
\gamma R^{ab}{}_{cd}R^{cd}{}_{ae})\xi^e
\end{eqnarray}
If this entire expression is to vanish by virtue of the equations of motion (\ref{gbequation}), then the terms proportional to $\nabla^a\xi^b$ must vanish identically.  
 Recall that the Riemann tensor satisfies the Bianchi identity $\nabla_{[a}R_{bc]de}=0$.   Contracting once and then twice with the inverse metric gives respectively the relations 
$\nabla_d R^d{}_{abc} - 2\nabla_{[b}R_{c]e}=0$ and $2\nabla_aR^a{}_b-\nabla_bR=0$.  It follows that the necessary cancelation of terms in  (\ref{divergence}) will occur if the coefficients in the boundary integrand are related according to integrand must be related according to $\alpha=-\beta/2 = \gamma$.  
Fixing the overall normalization of the boundary term by taking $\alpha=2$, one then has the result
\begin{equation}
Q =  \int  dS_a \calr^{(2)a}{}_b\,  \xi^b=0, \qquad .
\end{equation}
This Komar integral relation is clearly analogous to the result in Einstein gravity.

We now ask whether the Komar integral relations of pure Einstein and pure Gauss-Bonnet gravities can be generalized to all pure Lovelock theories?  In order to see how this will work, we can write the Komar boundary integrands, renamed $B^{(1)ab}$ for Einstein gravity and $B^{(2)ab}$ for Gauss-Bonnet gravity, in the more manifestly Lovelock form
\begin{equation}
B^{(1)ab} = \half\delta^{ab}_{cd}\nabla^c\xi^d,\qquad
B^{(2)ab} = \half\delta^{abcd}_{efgh} (\nabla^e\xi^f) R_{cd}{}^{gh}.
\end{equation}
The forms of these terms suggest that for $k$th order pure Lovelock gravity ({\it i.e.} with Lagrangian 
$\call=c\call_k$) one should try the boundary term
\begin{equation}\label{pureboundary}
B^{(k)ab}= {k\over 2^k}\delta^{abc_1d_1\dots c_{k-1}d_{k-1}}_{efg_1h_1\dots g_{k-1}h_{k-1}}
(\nabla^e\xi^f)R_{c_1d_1}{}^{g_1h_1}\cdots R_{c_{k-1}d_{k-1}}{}^{g_{k-1}h_{k-1}} .
\end{equation}

Now consider the computation of $\nabla_a\bkab$.  Potential contributions that come from the derivative acting on one of the curvature tensors all vanish due to the Bianchi identity 
$\nabla_{[a}R_{bc]}{}^{de}=0$ together with the antisymmetry of the $\delta$ symbol.  One is left with a single term that has two derivatives acting on the Killing vector.  After using the relation 
$\nabla_a\nabla_b\xi_c =-R_{bca}{}^d\xi_d$, one has $k$ powers of the curvature tensor contracted with the antisymmetric $\delta$ symbol and further manipulations lead to the result
\begin{equation}
\nabla_a\bkab = -\calr^{(k)b}{}_c\xi^c .
\end{equation}
Since the pure Lovelock equations of motion can be put in the form $\rkab=0$, we then have 
$\nabla_a\bkab=0$, and therefore the Komar integral relation
\begin{equation}
Q = \int_{\partial\Sigma}dS_{ab}\bkab =0.
\end{equation}

The Komar boundary integrands given above for pure Lovelock gravities are the duals of the Noether charges derived for these theories via the general methods of references \cite{Wald:1993nt,Iyer:1994ys}.  This relation will no longer hold for the Komar boundary integrands for general Lovelock theories.   The volume integrand for the Noether charges of  \cite{Wald:1993nt,Iyer:1994ys} is given by the interior product of the Killing vector with the Lagrangian, regarded as a $D$-form.  As noted in section (\ref{lovelocksection}), the Lagrangian  vanishes on solutions only for pure Lovelock gravities.

\section{Example: black holes in pure Lovelock gravity}\label{purebh}

One can check that the Komar boundary terms given in section \ref{purelovelock}  for pure Lovelock gravities give sensible results.  The static, spherically symmetric solutions of pure Lovelock gravity theories have been studied in reference \cite{Crisostomo:2000bb} and have the form
\begin{equation}\label{purestatic}
ds^2 = -f dt^2 + {dr^2\over f} +r^2 d\Omega_{D-2}^2,  \qquad f = 1 - \alpha/r^{{D-(2k+1)\over k}}
\end{equation}
where $\alpha$ is a dimensionful constant and it is assumed that $D\ge 2k+1$.  We write the metric on the unit  $(D-2)$-sphere in terms of coordinates $x^i$ as $d\Omega_{D-2}^2 = \gamma_{ij}dx^i dx^j$.  For our computation of the Komar boundary term, we will need the curvature component
\begin{equation}
R_{ij}{}^{kl} = {1\over r^2}(1-f)\delta^{kl}_{ij}
\end{equation}
We will also need the nonzero components of $\nabla^a \xi^b$ for the time translation Killing vector $\xi^a=(\partial/\partial t)^a$, which are given by $\nabla^r \xi^t = - \nabla^t \xi^r = \partial_r f/2$.

For $k=1$, {\it i.e.} pure Einstein gravity,  equation (\ref{purestatic}) reduces to the Schwarzschild solution and the metric function $f$ has the falloff at large $r$ characteristic of standard asymptotically flat boundary conditions.  However, for  $k\ge 2$ the metric function $f$ falls off less rapidly.  Nevertheless, it is argued in \cite{Crisostomo:2000bb} that a reasonable definition of the ADM mass  based on an analogue of the canonical formalism of  reference \cite{Regge:1974zd}  yields
\begin{equation}\label{adm}
M = c\alpha^k
\end{equation}
where $c$ is a numerical coefficient \cite{endnote0} (see also {\it e.g.} references 
\cite{Deser:2002rt,Deser:2002jk} for discussions of the definition of energy in 
higher derivative gravity theories).
In order to say whether, or not, the Komar boundary term in pure Lovelock gravity is sensible, we will ask whether the Komar boundary term evaluated at infinity for the time translation Killing vector, the analogue of the Komar mass, is finite, non-zero and also proportional to $\alpha^k$ for the spacetimes (\ref{purestatic}).

The Komar boundary integral over a sphere at spatial infinity is given by 
$Q_{\infty} = \int_\infty dS_{rt} B^{rt} $ with $dS_{rt} = {1\over 2}r^{D-2}d\Omega_{D-2}$.  Given that the only nonzero components of $\nabla^a \xi^b$ are $\nabla^r \xi^t = - \nabla^t \xi^r$, one finds that
\begin{eqnarray}
B^{rt} = {k\over 2^{k-1}}( \nabla^r \xi^t )\,  \delta^{i_1j_1\dots i_{p-1}j_{p-1}}_{k_1l_1\dots k_{p-1}l_{p-1}}\, 
R_{i_1j_1}{}^{k_1l_1}\cdots R_{i_{p-1}j_{p-1}}{}^{k_{p-1}l_{p-1}}
\end{eqnarray}
For the form of the metric function $f$ in (\ref{purestatic}), we have%
\begin{equation}
\nabla^r\xi^t = {D-(2k+1)\over k}\, \alpha\, r^{-{D-(2k+1)\over k}-1},\qquad
R_{ij}^{kl} = \alpha\, r^{-{D-(2k+1)\over k}-2}\delta_{ij}^{kl}
\end{equation}
Putting all these ingredients together, one finds that 
\begin{equation}
Q_\infty = \tilde c\alpha^k,
\end{equation}
where $\tilde c$ is another numerical factor .  Comparing with (\ref{adm}), we see that the Komar boundary term is finite, non-zero and proportional to the ADM mass for these spacetimes.

\section{Lower derivative gravity: $\Lambda\ne 0$?}\label{cosmoconstant}

Given that pure Lovelock gravity theories have Komar integral relations, it is now natural to ask whether   general Lovelock gravity theories do as well?  However, before addressing this question in general, recall that Lovelock gravity also has a contribution at zeroth order in curvature.  The simplest example of a more general Lovelock theory is therefore Einstein gravity with a non-zero cosmological constant.
Before proceeding to higher orders in curvature, we need to know whether Komar integral relations exist in lower curvature gravity, {\it i.e.} with $\Lambda\neq 0$.

This is clearly necessary in order for a result in general Lovelock gravity to hold.  However, it is of even more interest in its own right, given that  Einstein gravity with $\Lambda\neq 0$ may  well describe our universe.  Still, we will see that having a Lovelock perspective on this question is  useful in that it leads to a natural construction, that one might not otherwise have been led to consider.

Let us return to the Komar integral construction for Einstein gravity.
Clearly the volume integral (\ref{volumecharge}) does not vanish with $\Lambda\neq 0$, and hence we will no longer have a Komar integral relation.  This situation was considered some years ago in reference \cite{Magnon:1985sc}, where the main concern was to have a well-defined method of computing a finite Komar mass in asymptotically AdS spacetimes.  If one naively tries to evaluate the Komar mass in this case by computing the boundary integral (\ref{komarcharge}) at infinity, the result is infinite, reflecting the infinite volume contribution from $\Lambda\neq 0$.  The prescription of \cite{Magnon:1985sc} for obtaining a finite Komar mass involves regularizing the computation and making a background subtraction that becomes infinite as the regularization is removed.  The Komar mass computed in this way  for AdS spacetime itself then vanishes.  This background subtraction method has been used recently to compute the conserved quantities in higher dimensional Kerr-AdS spacetimes (see \textit{e.g.} reference \cite{Gibbons:2004ai}).
We will see that applying a Lovelock perspective to this problem yields an alternative construction.

Accordingly, let us ask whether it is possible to find a new Komar boundary integrand, such that the Einstein field equations with $\Lambda\neq 0$ make the corresponding volume integrand vanish?  
The Lovelock perspective suggests that we work out what sort of term is necessary by counting derivatives.  In Einstein gravity with $\Lambda=0$, which is quadratic in derivatives of the spacetime metric, the boundary term (\ref{komarcharge}) has one derivative acting on the Killing vector.
In the last section, we saw that in order to accommodate  the four derivative interactions of quadratic gravity, we had to have a Komar boundary integral which was cubic in derivatives, one derivative acting on the Killing vector and two in the curvature tensor.
A similar result held for all the pure Lovelock theories, $2k$-derivatives in the Lagrangian, required a Komar boundary term with $2k-1$ derivatives.  The cosmological constant term in the Lagrangian has $0$ derivatives.  Therefore, we should look for a term to add to the boundary integral that in some appropriate sense has $-1$ derivatives.

How can we make sense of a $-1$ derivative boundary term? Recall that a Killing vector satisfies $\nabla_a\xi^a=0$.  It is therefore possible, at least locally, to write the Killing vector in terms of a rank $2$ antisymmetric  potential $\omega^{ab}=\omega^{[ab]}$ as 
\begin{equation}\label{minusone}
\xi^b = \nabla_a\omega^{ab}.
\end{equation}
Note that $\omega^{ab}$ is not determined uniquely.  One can always add to it a term of the form 
$\nabla_c \lambda^{abc}$, corresponding to an exact form, or a term $\tilde\omega^{ab}$ that satisfies 
$\nabla_a\tilde\omega^{ab}=0$, corresponding to non-trivial cohomology, without affect the relation (\ref{minusone}).  Addition of an exact form will not affect the Komar boundary integral defined below.  We will see in the example in section (\ref{schwarz-ads}) that it is in practise not difficult to exclude the potential contribution from non-trivial cohomology.

The antisymmetric tensor $\omega^{ab}$ fits nicely into the formalism.
The Einstein equations with $\Lambda\ne 0$ are given by
$R_{ab} = 2\Lambda g_{ab}/(D-2)$.
%If we plug this into the Komar volume integral we get 
%%
%\begin{equation}\label{cosmokomar}
%\int_{\partial\Sigma} dS_{ab}\nabla^a\xi^b =  {2\over D-2}\Lambda\int_\Sigma dS_a\, \xi^a
%\end{equation}
%%
%Plugging equation (\ref{minusone})  into equation (\ref{cosmokomar}) gives
%%
%\begin{equation}
%\int_{\partial\Sigma} dS_{ab}\nabla^a\xi^b  =  {2\Lambda\over D-2}\int_{\Sigma} dS_{b}\nabla_a\omega^{ab}
%= - {2\Lambda\over D-2}\int_{\partial\Sigma} dS_{ab}\, \omega^{ab}
%\end{equation}
%Moving the boundary integral from the right hand side to the left, we then have a relation of the desired form, namely $Q=0$, with
It is straightforward to check that the new Komar boundary integral
\begin{equation}\label{lambdakomar}
Q = \int_{\partial\Sigma} dS_{ab}\left( \nabla^a\xi^b + {2\Lambda\over D-2}\omega^{ab}\right).
\end{equation}
satisfies $Q=0$.
Writing the Killing vector $\xi^a$ in terms of the antisymmetric potential $\omega^{ab}$ achieves the goal of acting with $-1$ derivatives on the Killing vector.

\section{Example: Schwarzschild-AdS}\label{schwarz-ads}
This new formulation for the Komar mass with $\Lambda\neq 0$ can be checked by computing the Komar mass for Schwarzschild-AdS spacetimes, 
\begin{equation}
ds^2 = -f dt^2 +{dr^2\over f} +r^2 d\Omega_{D-2}^2,\qquad 
f = 1 - {16\pi G M\over (D-3)\Omega_{D-2}r^{D-3}} 
- {2\Lambda\over (D-1)(D-2)}r^2
\end{equation}
where $M$ is the ADM mass as determined by {\it e.g.} by the methods of \cite{Abbott:1981ff}.  The Komar mass is equal, up to a normalization factor, to the Komar boundary term  evaluated at infinity for   $\xi^a=(\partial/\partial t)^a$.  

The quantity $\omega^{ab}$ appearing in the boundary term satisfies equation (\ref{minusone}).  We look for a spherically symmetric solution,  which reduces the problem to finding $\omega^{rt}$ such that 
${1\over r^{D-2}}\partial_r(r^{D-2} \omega^{rt}) = 1$.
This is solved by
\begin{equation}
\omega^{rt} = {r\over D-1} + {\eta\over r^{D-2}}.
\end{equation}
The second term is dual to the volume form on the $D-2$ sphere.  Its integral over a sphere of constant radius is independent of the radius.  Recalling that $\Sigma$ is a spatial slice with an inner boundary at the horizon, it is clear that this second term comes from the nontrivial cohomology of $\Sigma$.  We can exclude this contribution, which in any case would cancel out from a Smarr formula,  by setting $\eta=0$.   

Near infinity on a spatial slice, the volume element on a sphere is given by 
$dS_{rt}=-dS_{tr}=(1/2) r^{D-2}d\Omega_{D-2}$.
The derivative of the Killing vector is given by
\begin{equation}\label{twist}
\nabla_r\xi_t = -{1\over 2}\partial_r f 
= -{8\pi G M\over \Omega_{D-2} r^{D-2}} + {2\Lambda r\over (D-1)(D-2)}
\end{equation}
The boundary integral at infinity, $Q_\infty$, is then found to be 
\begin{equation}
Q_\infty =  2 \int_{\partial\Sigma_\infty} dS_{rt} \left( \nabla^r\xi^t + {2\Lambda\over D-2}\omega^{rt}\right)
=8\pi GM.
\end{equation}
The divergent contribution of the second term in (\ref{twist})  is exactly cancelled by the contribution of the antisymmetric tensor potential.
The infinite background subtraction of the prescription of \cite{Magnon:1985sc} is effectively carried out by cancellations within the boundary integrand itself \cite{endnote3}.

\section{Komar integrals in general Lovelock gravity}\label{generalsection}

A Komar integral relation was found in section (\ref{cosmoconstant}) for the simplest non-pure Lovelock theory, Einstein gravity with $\Lambda\neq 0$.  This construction required adding a new contribution $\omega^{ab}$ to the boundary term, whose existence followed from the divergence-free property of the Killing vector $\xi^a$.  The Komar construction in general Lovelock gravity will require additional terms of a similar nature.

A current divergence free current $\jka$ can be associated with each of the Lovelock terms by the definition
\begin{equation}
\jka=\gkab\xi^b.
\end{equation}
Therefore at least locally one can write
\begin{equation}
\jka =\nabla_b\omega^{(k)ba} 
\end{equation}
with $\omega^{(k)ab}$ antisymmetric \cite{endnote1}.  A candidate Komar boundary term for general Lovelock gravity includes both the forms $\bkab$ and $\okab$ with arbitrary coefficients
\begin{equation}
B^{ab} = \sum_{k\le\kbar}\left(d_k\bkab +e_k\okab\right).
\end{equation}
Its divergence is given by
\begin{equation}\label{bigdivergence}
\nabla_a\bkab = \sum_{k\le\kbar} (-d_k\calr^{\pk b}{}_c + e_k\calg^{\pk b}{}_c)\xi^c
\end{equation}
The coefficients $d_k$ and $e_k$ are determined by requiring  that $\nabla_aB^{ab}=0$ by the equations of motion.  There turn out to be a number of ways to do this.  One way is to simply set the coefficients $d_k=0$  and to take all the coefficients $e_k$ equal.  The right hand side of (\ref{bigdivergence}) is then simply proportional to the equations of motion in their original form $\sum_{k\le\kbar} c_k\calg^{\pk a}{}_b=0$.  However, for pure Einstein gravity this choice would not give the Komar boundary term.  

It is also necessary to require that for $\kbar=1$ we recover the Komar boundary term of section  (\ref{cosmoconstant}).  The following choice satisfies this criterion and leads to a unique solution for the remaining coefficients.   Take $e_{\kbar}=0$ and fix the overall normalization of the boundary term by taking $d_\kbar = c_\kbar$.  Equation (\ref{bigdivergence}) can then be rewritten as 
\begin{equation}\label{generaldiv}
\nabla_a\bkab =
-c_\kbar\calr^{(\kbar) b}{}_c\xi^c + 
\sum_{k<\kbar}\left( (e_k - d_k) \calr^{\pk b}{}_c -\half e_kg^b{}_c\callk\right)\xi^c,
\end{equation}
One can now solve the equations of motion for $\calr^{(\kbar)a}{}_b$ in terms of the other quantities appearing in (\ref{generaldiv}) to get 
\begin{equation}
c_\kbar  \calr^{(\kbar)a}{}_b = -\sum_{k<\kbar} c_k\left(
\rkab + ({\kbar-k\over D-2\kbar})\gab\callk\right).
\end{equation}
Plugging this into equation (\ref{generaldiv}), and requiring that the coefficients of the quanties $\rkab$ and  $\call_k$ vanish then fixes the coefficients $d_k$ and $e_k$ with $k<\kbar$.  The resulting Komar boundary integrand for
%%
%\begin{equation}
%d_k = ({D-2k\over D-2\bar k})c_k,\qquad e_k = 2 ({\kbar -k\over D-2\kbar}) c_k,\qquad  k<\kbar
%\end{equation}
%%
a general Lovelock theory is then given by
\begin{equation}\label{bigboundary}
B^{ab} = c_\kbar B^{(\kbar)ab} + \sum_{k<\kbar}
c_k\left( ({D-2k\over D-2\kbar})\bkab +2({\kbar-k\over D-2\kbar})\okab\right).
\end{equation}

The starting choice of $d_\kbar=c_\kbar$ and $e_\kbar=0$ is not entirely unique.  The criteria $\nabla_aB^{ab}=0$ can also be satisfied by picking a different value $\hat k$ in the range $1<\hat k\le \kbar$ and setting $d_\khat=c_\khat$ and $e_\khat=0$.  The resulting boundary term would have the same form as (\ref{bigboundary}) but with $\khat$ replacing $\kbar$.  Always making the  choice $\khat =1$, independent of the value of $\kbar$,  would then also satisfy the additional criteria that for $\kbar=1$ one gets back the correct Komar boundary term.  

The choice made above privileges the highest order term in the theory, while the alternate choice privileges the Einstein term.   One way to justify the choice of $\khat=\kbar$  is to note that given a  choice of $\khat$, the calculation of the remaining coefficients in (\ref{bigboundary}) requires that $c_\khat\neq 0$, while the other coefficients $c_k$ in the Lagrangian can be freely varied.  Keeping $c_\kbar\neq 0$ fixes the behavior of the theory at large curvature, an important aspect of the theory, determining {\it e.g.} the number of constant curvature vacua.
Keeping $c_\khat\neq 0$ for some $\khat<\kbar$, on the other hand, is more like fixing a detail of the theory, rather than its general character.

One may be interested in the nonvacuum case as well.  The equations of motion are then given by
$\calg^a{}_b = 8\pi T^a{}_b$
where $\calg^a{}_b\equiv\sum_{k\le\kbar}\gkab$.  The divergence of the boundary term (\ref{bigboundary}) is then given by
\begin{equation}\label{nonvacuum}
\nabla_a B^{ab} = -8\pi (T^b{}_c - {1\over D-2\kbar}g^v{}_c T^a{}_a)\xi^c
\end{equation}
In this case the Komar integral relation will involve a non-zero volume term as well \cite{endnote2}.

\section{Example: Gauss-Bonnet gravity}\label{GBsection}

The full expression (\ref{bigboundary}) for the Komar boundary term in Lovelock gravity is fairly complicated, and it is natural to want to check that it gives sensible results.  This was done already in section (\ref{cosmoconstant}) for the case $\kbar=1$, Einstein gravity with $\Lambda\neq 0$.  In this section, we work out an example that includes higher derivative interactions as will.  Consider a general Gauss-Bonnet gravity theory, described by the Lagrangian
\begin{equation}
\call = c_2\call_2+c_1\call_1 + c_0\call_0.
\end{equation}
Static, vacuum black holes in this theory have been studied in references 
\cite{Boulware:1985wk,Wheeler:1985nh,Wheeler:1985qd,Wiltshire:1985us,Wiltshire:1988uq,Whitt:1988ax,Cai:2001dz,Cai:2003gr}.  These have the form
\begin{equation}
ds^2 = -f(r)dt^2 +{dr^2\over f(r)} +r^2d\Omega_{D-2}^2
\end{equation}
where the metric function $f(r)$ can be found from the $tt$ component of the equations of motion, which requires that 
\begin{equation}
\partial_r\left[
r^{D-1}\left( \hat c_2{(1-f)^2\over r^4}+\hat c_1{(1-f)\over r^2} + \hat c_0\right)\right]=0
\end{equation}
where $\hat c_0=c_0$, $\hat c_1=(D-1)(D-2)c_1$ and $\hat c_2=(D-1)(D-2)(D-3)(D-4)c_2$.

The metric function $f$ is then easily found by writing $F=1-f$ and solving the equation
\begin{equation}\label{polynomial}
P[F] = {\lambda\over r^{D-1}},\qquad P[F] =\hat c_2{F^2\over r^4} +\hat c_1{F\over r^2} +\hat c_0
\end{equation}
with $\lambda$ an arbitrary constant.  The explicit solutions are given by
\begin{equation}
f(r) = 1 + {r^2\hat c_1\over 2\hat c_2}\left( -\hat c_1\pm
\sqrt{\hat c_1^2-4(\hat c_0-{\lambda\over r^{D-1}})\hat c_2}\right)
\end{equation}
Assuming that $\hat c_1^2 - 4\hat c_0\hat c_2>0$, the solutions are generally asymptotically (A)dS, with 
asymptotically flat solutions only in the $+$ branch for $c_0=0$.  The parameter $\lambda$ is proportional to the mass of the spacetime \cite{Cai:2001dz}.

Let us now evaluate the Komar boundary integral for a sphere of constant radius on a constant time slice.  If the boundary term (\ref{bigboundary}) is sensible, then we expect that in the limit of large radius, the Komar integral should be finite and proportional to the gravitational mass, {\it i.e.}  to the parameter $\lambda$.  However, because $\nabla_aB^{ab}=0$, the integral should be independent of the radius of the sphere.  Let us see how this comes about.

Plugging $\kbar =2$ into the expression (\ref{bigboundary}) for the boundary term, we find
\begin{equation}
B^{ab} = c_2 B^{(2)ab} + {D-2\over D-4} c_1 B^{(1)ab} + {2\over D-4} c_1\omega^{(1)ab}
+{4\over D-4}c_0\omega^{(0)ab}.
\end{equation}
For the integral over the sphere, we need only the component $B^{rt}$.  The necessary ingredients are found to be
\begin{equation}
B^{(2)rt} = {(D-2)(D-3)\over r^2}(1-f)\partial_r f,\qquad B^{(1)rt}=\half\partial_r f, 
\end{equation}
\begin{equation}
\omega^{(1)rt} = - {D-2\over 2r}(1-f),\qquad \omega^{(0)rt} = -{r\over 2 (D-1)}.
\end{equation}
Assembling these, one finds that $B^{rt}$ can be expressed in terms of the function $P[F]$ given above in equation (\ref{polynomial}),
\begin{equation}
B^{rt} = - {r^2\over (D-1)(D-4)}\left\{\half\partial_rP[F] + {2\over r}P[F]\right\} = 
{(D-5)\lambda\over 2(D-1)(D-4) r^{D-2}}.
\end{equation}
Integrating over a sphere of radius $R$ one finds
\begin{equation}
Q_R = {(D-5)\Omega_{D-2}\lambda\over 2(D-1)(D-4)}
\end{equation}
which is independent of the radius of the sphere and proportional to the gravitational mass.
Recall that in section (\ref{cosmoconstant}),  the additional term $\omega^{ab}$ added to the Komar boundary integral of  pure Einstein gravity served to cancel out the infinite contribution coming from the cosmological constant.  The boundary term in Gauss-Bonnet gravity makes this work out for both of the two possible asymptotic constant curvature vacua.    
Note that the metric function for a  black hole in a general Lovelock theory is given by a solution to a polynomial equation similar to (\ref{polynomial}) \cite{Wheeler:1985qd} (see also \cite{Myers:1988ze}).  It is natural to expect that the Komar boundary term for a constant radius sphere in such a spacetime can similarly be written in terms of this polynomial.    
\section{Conclusions}\label{conclusions}

Through a series of steps we have developed a generalization of the Komar integral construction that holds in a general Lovelock gravity theory and checked that it gives sensible results.  There are a number of further directions that would be interesting to pursue.  

As noted in the introduction, the Komar construction in Einstein gravity is used to derive the Smarr formula (\ref{smarr}).  One should also be able to do this for Lovelock theories \cite{endnote4}.  The terms $\bkab$ include $k-1$ powers of the curvature tensor and appear to have the right forms to reproduce the Lovelock contributions to black hole entropy \cite{Jacobson:1993xs} when evaluated on the horizon.  The terms $\okab$, however, are new and their interpretation in black hole thermodynamics will need to be clarified.  This is already the case for the $\omega^{(0)ab}$ contribution in Einstein gravity with a nonzero cosmological constant.  Smarr formulas are generally related to the first law of black hole thermodynamics via a scaling argument (see {\it e.g.} references \cite{Chowdhury:2006qn,Kastor:2008wd})).  It was suggested in \cite{Gibbons:2004ai} that the Smarr formula with $\Lambda\ne 0$ may be connected to a first law that includes variations in $\Lambda$.  Such an approach was taken in reference \cite{Wang:2006bn}, and the Smarr formula found there by integrating the first law includes an additional term proportional to $\Lambda$.  Possibly the other terms in the Smarr formula, coming from integrating the forms $\okab$ on the horizon, have a similar origin in a first law that includes variations in the coefficients of the subleading terms in the Lovelock Lagrangian.  

Finally, having established that the Komar construction holds in the Lovelock subclass, it will be interesting to return to the question of whether Komar integral relations exist in more general higher derivative gravity theories.  In this case, the equations of motion will depend on derivatives of the curvature tensor as well.  For quadratic curvature theories, additional boundary terms such as $(\nabla^{[a}R)\xi^{b]}$ and $(\nabla^{[a}R^{b]}_c)\xi^c$ can arise.  This analysis is left for future work.

\subsection*{Acknowledgements}
The author thanks Alex Maloney and Jennie Traschen for helpful discussions.  This work was supported in part by NSF grant PHY-0555304.


\begin{thebibliography}{99}
\bibitem{endnote0}
  In this and subsequent examples, we will be concerned primarily with the structure of the results and will not try to determine the overall normalization factors necessary to get precise agreement with other definitions of the mass.
\bibitem{endnote1}
Note that $\omega^{(0)ab}$ defined in this way is related to $\omega^{ab}$ from section (\ref{cosmoconstant}) by a factor of 
$-(1/2)$ because $\calg^{(0)a}{}_b = -(1/2)g^a{}_b$.
\end{thebibliography}
 \end{document}